\documentclass[12pt,aps,nofootinbib]{revtex4}

\usepackage{graphicx,amsmath,amssymb}
\usepackage{bm}

\begin{document}


\title{Instability of the Rayleigh-Jeans spectrum \\  
in weak wave turbulence theory}

\author{Miguel Escobedo}
\affiliation
      {%
       Departamento de Matem\'aticas, 
       Universidad del Pa\'\i s Vasco, 
       Apartado 644, E-48080 Bilbao, Spain
       }%
\author{Manuel A.Valle}
\affiliation
      {%
       Departamento de F\'\i sica Te\'orica, 
       Universidad del Pa\'\i s Vasco, 
       Apartado 644, E-48080 Bilbao, Spain
       }%

\date{\today}

\begin{abstract}
We study the four-wave kinetic equation of weak turbulence linearized around the Rayleigh-Jeans spectrum 
when the collision integral is associated with short-range interactions between non-relativistic bosonic quasiparticles. 
The technique used for the analysis of the stability  is based on the properties of the Mellin transform of the kernel in the integral equation. 
We find that any perturbation of the Rayleigh-Jeans distribution evolves towards low momentum scales in such a form that, 
when $t \rightarrow \infty$, all the particles occupy  a sphere of radius arbitrary small.  
\end{abstract}



\maketitle


\section{\label{sec:intro} Introduction}

Since early work that has addressed the kinetics of Bose-Einstein condensation even before the  
experimental realization of BEC in weakly interacting atomic  gases~\cite{Levich,Snoke,Stoof1,Kagan0,Semikoz1},  
the description of the growth of the condensate has been the subject of 
considerable attention~\cite{Zoller,Stoof2,Kagan,Semikoz2,Zaremba}. 
At the first stage of evolution, a description based on the Boltzmann kinetic equation is adequate. 
In the homogeneous case, it has the form    
\begin{eqnarray}
\partial_t n(\bm{p}_1, t) &=& \frac{1}{2} \int_{\bm{p}_2 \bm{p}_3 \bm{p}_4} |\mathcal{M}|^2
(2 \pi)^3 \delta(\bm{p}_1+\bm{p}_2 -\bm{p}_3 -\bm{p}_4) 
 2 \pi \delta(\omega_1+ \omega_2-\omega_3-\omega_4)    \nonumber  \\
 && \times \left[ (1+n_1)(1+n_2) n_3 n_4 -  n_1 n_2(1+n_3)(1+ n_4) \right], 
\end{eqnarray}
where we are using $\int_{\bm{p}}$ to denote momentum integration,
\begin{equation}
\int_{\bm{p}} \cdots \equiv \int \frac{d^3 \bm{p}}{(2\pi)^3} \cdots .
\end{equation}
Here $n(\bm{p},t)$ is the distribution function, the average density of particles with momentum $\bm{p}$ at time $t$.  
The square-amplitude is  taken as $ |\mathcal{M}|^2 =  (8\pi a m^{-1})^2$, with $a$ the scattering length 
parameterizing the Fermi pseudopotential  $V(\bm{r}_1-\bm{r}_2) =  4\pi a m^{-1} \delta(\bm{r}_1-\bm{r}_2)$ 
appropriate to low-energy scattering by a interaction of finite range $R\gg a$. 
The dispersion law is $\omega(p) = \frac{p^2}{2 m}$. 

\begin{figure}
\centering
    \includegraphics{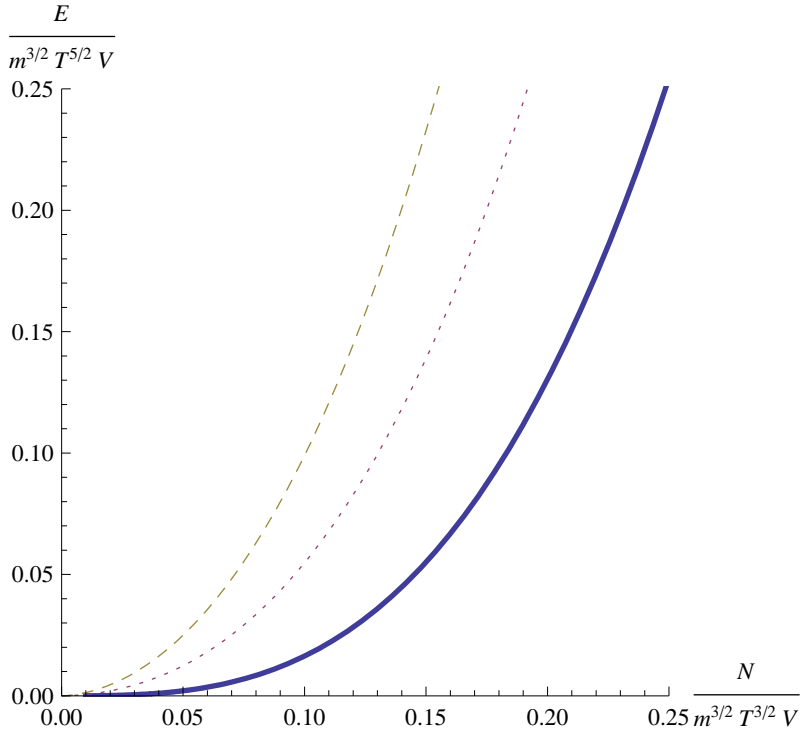}
    \caption{\label{fig:phase}(color online). Parameter space at equilibrium for large occupation numbers.  The curves are lines of $\mu/T$ constant, 
    obtained as parametric representations of the scaled quantities $N/V$ and $E/V$ in terms of the parameter $\Lambda$. 
     The thick line corresponds to $\mu=0$. Each point on this curve corresponds to a Rayleigh-Jeans distribution with a critical temperature $T_c=\pi^2 N/(m V \Lambda) = 6\pi^2 E/(V \Lambda^3)$ where $N/V$ and $E/V$ are the particle and energy density at the prescribed equilibrium. 
Each point in the region above the $\mu=0$ line is associated with a pair $(T,\mu<0)$ of the equilibrium distribution $n(p)=T/(\omega(p)-\mu))$.     
     The dotted and dashed curves correspond to $\mu/T=-0.1$ and $-0.4$ respectively. For a point below the $\mu=0$ line, the departure of its abcise from the curve corresponds to the particle density in the condensate and the energy density is given by its ordinate.} 
\end{figure}

Based on the assumption that the formation of a condensate is a low energy process where the occupation numbers obey 
the condition $n(t,\bm{p}) \gg 1$, it is customary to approximate the Boltzmann equation by the kinetic equation of 
wave turbulence~\cite{Newell, book} 
\begin{eqnarray}
\label{eq:kin4}
\partial_t n_1 &=&  \frac{1}{2} \left(\frac{8\pi a}{m}\right)^2 \int_{\bm{p}_2 \bm{p}_3 \bm{p}_4} 
(2 \pi)^3 \delta(\bm{p}_1+\bm{p}_2 -\bm{p}_3 -\bm{p}_4) 
2 \pi \delta(\omega_1+ \omega_2-\omega_3-\omega_4)   \nonumber \\
&&\times \left[(n_1+ n_2) n_3 n_4 -  n_1 n_2(n_3+ n_4) \right]. 
\end{eqnarray}
Using considerations of homogeneity~\cite{Balk2}, it is possible to obtain stationary power-law  solutions  $n = B p^{-\nu}$, 
for the pair of values $\nu = 3, 7/3$. These Kolmogorov spectra are conjugate to the obvious solutions for 
$\nu=2, 0$.  However, the results of several studies~\cite{Kagan0,Kagan,Semikoz1,Semikoz2,Lacaze}  which have addressed 
the question of the dynamics underlying Eq.~(\ref{eq:kin4})   
seem to exclude any significant role of the Kolmogorov spectra in the formation 
of a finite time singularity corresponding to a condensate.
Within the framework of kinetic theory, the dynamics before blow-up is rather described 
by a self-similar solution of the form $n(t,p)=\beta^{-1/2} \Phi((t_\ast-t)^{-\beta}/p^2)$, 
where $t_\ast$ is the time of blow-up and the parameters $\alpha$ and $\beta$ are related by $\beta=\alpha-1/2$  (for a review, see Ref.~\cite{Josserand}). 
The determination of the ratio $\alpha/\beta$ poses a very difficult nonlinear eigenvalue problem  which can not be solved by scaling arguments. 
The numerical observed value $\alpha/\beta \approx 1.234$ is significant different from the exponents $\nu/2 =3/2, 7/6, 1$ associated with Kolmogorov and equilibrium solutions.  
However, the classical Rayleigh-Jeans spectrum with zero chemical potential plays a special role as boundary in the appropriate parameter space between the regions of condensate and normal phases (see Fig.~\ref{fig:phase}). 
To understand its significance it can be of some interest to study in detail the stability properties of this equilibrium solution.

In this paper, we address the question of the linear stability  of the Rayleigh-Jeans spectrum ($\nu=2$) of (\ref{eq:kin4})  
whose perturbations could be related to processes driving particle transport towards the region of $\bm{p}=0$. 
Let us remark that if one tries to use the equilibrium $n = T/\omega(p)$ in the entire interval of time, one faces a serious difficulty. 
Consider a prescribed total particle density $n$ to be accommodated  by a singular distribution $n_0(T)(2\pi)^3 \delta(\bm{p})$ and  
an equilibrium Rayleigh-Jeans distribution through
\begin{equation}
n = n_0(T) + \int_0^\Lambda \frac{T}{p^2/(2 m)} \frac{4 \pi p^2 d p}{(2\pi)^3}  = n_0(T) +  \frac{m T}{\pi^2} \Lambda . 
\end{equation} 
For a given cut-off $\Lambda$, the second term grows with $T$, and when it reaches the prescribed total density $n$, the density of particles in the 
$\bm{p}=0$ state vanishes. The temperature above which $n_0=0$ is $T_c =  \pi^2 n/(m \Lambda)$. In this model, 
the particle density in the condensate is given by 
\begin{equation}
 n_0(T) = n \left(1 - T/T_c \right), \quad  T \leq T_c .
\end{equation}
From this fact, one would expect that the Rayleigh-Jeans spectrum can not be stable with respect to changes of temperature: a small change in $T$ in the neighborhood of $T_c$ may move the system toward a state where $n_0 \neq 0$ or not. 
Remarkably, 
this will be reflected by the fact that the linearized kinetic equation about  $n^0 = T/\omega(p)$ does not admit  
as a solution the function $A(\bm{p},t) \equiv \delta T(\bm{p},t)/T = \text{constant}$. 
This argument pointing out an instability does not apply to the equilibrium distribution function, $n \propto p^{-1}$, 
which arises as  the solution of the three-wave kinetic equation at the end 
of the condensation process~\cite{Nazarenko,Rica}. 
This comes from to the low frequency limit of the Bogoliubov dispersion law $\omega_B(p) = \sqrt{4\pi a p^2/m  + (p^2/(2m))^2}$. 
In that case, the distribution function $n$ corresponds to Goldstone-type excitations whose number is not conserved. 
Our focus will be on the regime in which the four-wave interactions are the most important. 
The transition from this stage to the three-wave regime  has been recently studied  within the two-dimensional Gross-Pitaevskii model in~\cite{Nazarenko2,Nazarenko3}. 
These authors have  shown  that this 
transition  involves an intermediate step where 
strong interactions between topological defects take place    
in a similar way to the Kibble-Zurek mechanism in continuous phase transitions.



It is important to point out that the evolution of an initial distribution function $n(\bm{p},t=0)$ 
in the regime of large occupation numbers  produces a change in the entropy density 
\begin{equation}
\frac{S}{V} = \int_{\bm{p}}  \left[ (1+n) \ln (1+n) - n \ln n \right]
\end{equation}
given by 
\begin{eqnarray}
\frac{d(S/V)}{dt}&=&  \frac{1}{4} \left(\frac{8\pi a}{m}\right)^2  
   \int_{\bm{p}_1\bm{p}_2 \bm{p}_3 \bm{p}_4} 
  (2 \pi)^3 \delta(\bm{p}_1+\bm{p}_2 -\bm{p}_3 -\bm{p}_4) 
  2 \pi \delta(\omega_1+ \omega_2-\omega_3-\omega_4)   \nonumber \\
  &&\times n_1 n_2 n_3 n_4 \left(\frac{1}{n_1} +\frac{1}{n_2}-\frac{1}{n_3}-\frac{1}{n_4}\right)^2, 
\end{eqnarray}
which  only vanishes for distribution functions with the shape of Rayleigh-Jeans. 
This result and the conservation laws imply that 
the final state of the evolution is consistent with the presence of a regular Rayleigh-Jeans distribution  with or without a singular distribution, depending on whether the particle and energy densities in the initial distribution are below or above their critical values (see Fig.~\ref{fig:phase}).
It follows that a perturbed Rayleigh-Jeans distribution 
$T/\omega(p)+\delta n(\bm{p},t=0)$ can not evolve to the same  
$n^0=T/\omega(p)$ unless the energy $\delta(E/V)$ and the particle number $\delta(N/V)$ of $\delta n(\bm{p},t=0)$ vanish.

Although the theory of wave turbulence has been used  in a variety of systems, 
ranging from ocean or capillary waves to plasma waves or elastic waves 
(see, for example,~\cite{Connaughton} for a short review), 
specific studies on the stability of the  stationary spectra are rather scarce. 
In this respect, another objective  of this work is to present  a specific  new application 
where the techniques of Ref.~\cite{book}  are used. 

This paper is organized as follows. In Sec.~\ref{sec:lin} we derive the linearized kinetic equation about the Rayleigh-Jeans spectrum. 
In Sec.~\ref{sec:stb}, we introduce the Mellin function, and present some of its properties.  
The character of the evolution of the perturbation is considered as $p \rightarrow 0, \infty$ and 
as $t\rightarrow \infty$. 
We also compute the leading asymptotics as $t\rightarrow \infty$ of the perturbed particle number  which is contained in the 
interval $0 \leq p \leq \Lambda$. Section~\ref{sec:end} gives our conclusions. 
Details of some  calculations regarding the reduction of the integral equation to the Fredholm form 
and the evaluation of the Mellin function can be found in a pair of appendices.

\section{\label{sec:lin} The linearized kinetic equation} 

Here we derive the linearized kinetic equation about the Rayleigh-Jeans spectrum
\begin{equation}
n^0(p) = T \omega(p)^{-1} , 
\end{equation}
where it is understood that the temperature is at the critical value $T=T_c =  \pi^2 n/(m \Lambda)$ in this model. 
We shall closely follow the conventions of~\cite{book}. The dimensionless quantity 
$A(\bm{p},t)$ represents the relative change in the particle number and may be either positive or negative.  
Substituting the expression 
\begin{equation}
 n(\bm{p},t) = n^0(p) (1 + A(\bm{p},t))
\end{equation} 
into (\ref{eq:kin4}) and retaining the terms of $O(A)$  gives an integral equation for $A(\bm{p},t)$,
\begin{equation}
\partial_t A = L[A], 
\end{equation}
with the result
\begin{eqnarray}
\label{eq:kinlin}
L[A] &=&  \frac{1}{2} \left(\frac{8\pi a}{m}\right)^2  \int_{\bm{p}_2 \bm{p}_3 \bm{p}_4} 
  (2 \pi)^3 \delta(\bm{p}_1+\bm{p}_2 -\bm{p}_3 -\bm{p}_4) 
  2 \pi \delta(\omega_1+ \omega_2-\omega_3-\omega_4)   \nonumber \\
  &&\times \left\{\left[-n^0(p_2) n^0(p_3) - n^0(p_2) n^0(p_4) +  n^0(p_3) n^0(p_4) \right] A(\bm{p}_1,t) \right. \nonumber \\ 
&& \left.  -n^0(p_3) n^0(p_4) A(\bm{p}_2,t) + 2 \, n^0(p_2) n^0(p_4) A(\bm{p}_3,t) \right \}, 
\end{eqnarray}
where we have used the conservation of energy
\begin{equation}
\frac{1}{n^0(p_1)} + \frac{1}{n^0(p_2)} = \frac{1}{n^0(p_3)} +\frac{1}{n^0(p_4)} , 
\end{equation}
and the symmetry of the integrand under $\bm{p}_3 \leftrightarrow \bm{p}_4$ to duplicate the last term. 
As we shall see, one must be cautious using this rearrangement, 
which assumes regularity of the integrand. 
Here we require that the solution $A(\bm{p},t)$ falls off sufficiently rapid as $p \rightarrow \infty$ to permit 
this kind of reordering in the integration. In the present case the several integral terms of the operator $L$ 
containing the function $A(\bm{p},t)$ for different arguments $\bm{p}$ are well defined separately, 
and it is possible to write 
\begin{equation}
L[A] = N A(\bm{p},t) + \int U(\bm{p}, \bm{p}') A(\bm{p}',t) d^3 \bm{p}' , 
\end{equation} 
where the coefficient $N$ turns out to be a constant. 

Due to the conservation laws, one expects that the collision integral $L[A]$ vanishes for certain forms of $A$. One of them, which reflects conservation of particle number, is $A(p) \propto p^{-2}$. This collision invariant corresponds to the fact that the kinetic equation (\ref{eq:kin4}) admits the Rayleigh-Jeans spectrum with non-zero chemical potential as an exact solution 
\begin{equation}
n(p) = \frac{T}{\omega(p)-\mu} , 
\end{equation}
which implies that $A=\mu/\omega(p)$ is a zero mode of the linearized equation and 
\begin{equation}
 N = -\int \frac{p^2}{p'^2} U(\bm{p},\bm{p}') d^3 \bm{p}'
\end{equation}
would be true. In fact, this turns out to be the case, as we have explicitly checked 
in Eq.(\ref{eq:zm}) of the appendix. 
Since the linear momentum is conserved, a possible solution of Eq.~(\ref{eq:kin4}) is 
\begin{equation}
n(p) = \frac{T}{\omega(p)-\bm{v}\cdot\bm{p}} ,
\end{equation}
which would imply that $A \propto \bm{v}\cdot\bm{p}$ is also a zero mode of the operator $L$. 

Finally, from the conservation of the kinetic energy in collisions one would expect a zero mode $A = \textrm{constant}$ but, surprisingly, this does not occur, at least for the operator $L$ in the form (\ref{eq:kinlin}). Consider $L[1]$ as follows from Eq.~(\ref{eq:kinlin}):  
\begin{eqnarray}
L[1] &=& \frac{1}{2} \left(\frac{8\pi a}{m}\right)^2  \int_{\bm{p}_2 \bm{p}_3 \bm{p}_4} 
  (2 \pi)^3 \delta(\bm{p}_1+\bm{p}_2 -\bm{p}_3 -\bm{p}_4) 
  2 \pi \delta(\omega_1+ \omega_2-\omega_3-\omega_4)   \nonumber \\
  &&\times \left[-n^0(p_2) n^0(p_3)+ n^0(p_2) n^0(p_4)\right] .
\end{eqnarray}
It it were allowable to make the exchange $\bm{p}_3 \leftrightarrow \bm{p}_4$ in the second term, we would obtain the expected result $L[1]=0$. 
However, one finds that 
\begin{eqnarray}
\label{eq:asymp1}
\int_{\bm{p}_2 \bm{p}_4} \delta(\bm{p}_1+\bm{p}_2 -\bm{p}_3 -\bm{p}_4) 
  \delta(\omega_1+ \omega_2-\omega_3-\omega_4) n^0(p_2) n^0(p_4) \nonumber \\ 
  \qquad \sim -\frac{m^3 T^2}{2\pi^4} \frac{1}{p_3^3} 
  \left( \ln\frac{p_1}{p_3}-1 \right), \qquad p_3 \rightarrow \infty, 
\end{eqnarray}
and 
\begin{eqnarray}
\label{eq:asymp2}
\int_{\bm{p}_2 \bm{p}_4} \delta(\bm{p}_1+\bm{p}_2 -\bm{p}_3 -\bm{p}_4) 
  \delta(\omega_1+ \omega_2-\omega_3-\omega_4) n^0(p_2) n^0(p_3) \nonumber \\ 
  \qquad \sim -\frac{m^3 T^2}{2\pi^4} \frac{1}{p_2^3} 
  \left( \ln\frac{p_1}{p_2}-1 \right), \qquad p_2 \rightarrow \infty.  
\end{eqnarray}
These asymptotics give rise to leading divergences of the form $\ln^2(\Lambda/p_1)$ for the remanining integration over $\bm{p}_3$ and $\bm{p}_2$ respectively, being $\Lambda$ an ultraviolet cutoff. 
This points out that it is not legitimate to use the previous rearrangement. 
We can see, however, that there is a cancellation between both divergences with the result that $L[1]>0$. 
Using the notation in Appendix~\ref{sec:fredholm} we obtain 
\begin{equation}
\tau L[1] = \gamma_N + \int_0^\infty \mathcal{V}_0 \frac{d x}{x} = 
            \gamma_N + 2 G + \frac{3\pi^2}{8} \approx 0.8225.   
\end{equation}

It is convenient to expand the kernel $U(\bm{p},\bm{p}')$ in terms of Legendre polynomials of the 
angle $\theta_{\bm{p}\bm{p}'} = \hat{\bm{p} }\cdot \hat{\bm{p}'}$, 
\begin{equation}
U(\bm{p}, \bm{p}') = \sum_{l=0}^\infty  \mathcal{U}_l(p, p') P_l(\cos \theta_{\bm{p} \bm{p}'}), 
\end{equation}
and to expand~\cite{book} the perturbation $A$ in spherical harmonics relative to the orientation of $\bm{p}$,
\begin{equation}
A(\bm{p},t) = \sum_{l,m}  \mathcal{A}_{l m}(p,t) Y_l^m(\hat{\bm{p}}). 
\end{equation}
From the addition theorem for spherical harmonics 
\begin{equation}
 P_l(\cos \theta_{\bm{p} \bm{p}'}) = \frac{4 \pi}{2 l+1}\sum_{m=-l}^l Y_l^m(\hat{\bm{p}}) Y_l^m(\hat{\bm{p}'})^\ast , 
\end{equation} 
and the orthogonality relation ($d\Omega_{\bm{p}'}$ denotes the corresponding surface element of the three-dimensional unit sphere)
\begin{equation}
\int d\Omega_{\bm{p}'} Y_l^m(\hat{\bm{p}'}) Y_{l'}^{m'} (\hat{\bm{p}'})^\ast = \delta_{l l'} \delta_{m m'}, 
\end{equation}
a set of uncoupled equations emerges, each one of them labeled by $(l,m)$:  
\begin{equation}
\partial_t \mathcal{A}_{l m}(p,t) = N(p) \mathcal{A}_{l m}(p,t)+ \frac{4 \pi}{2 l+1}
\int_0^\infty \mathcal{U}_l(p,p') \mathcal{A}_{l m}(p',t) p'^2 d p' .
\end{equation} 
In the homogeneous case, the angular momentum projection $m$ may be suppresed because of isotropy, 
and it is enough to use  $l$ to label each irreducible perturbation. 

\begin{figure}
\centering
    \includegraphics{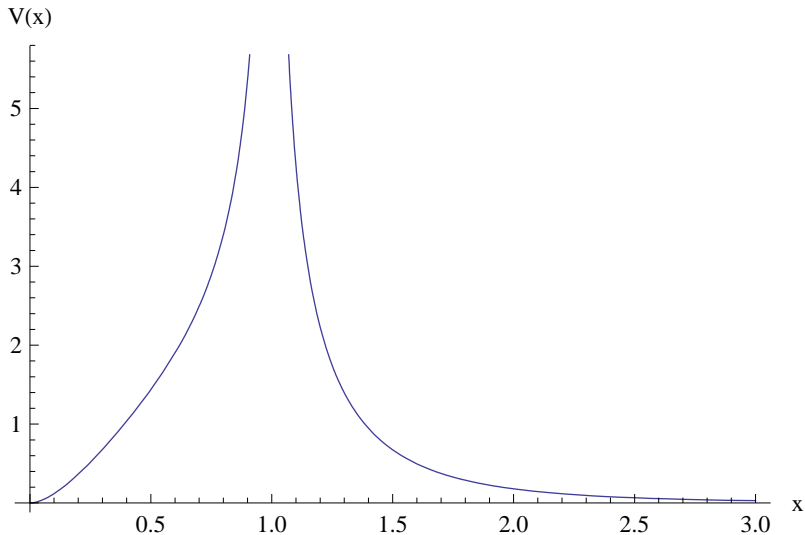}
    \caption{\label{fig:kernel}(color online).  A plot of the kernel $\mathcal{V}(x)$ for $0 \leq x \leq 3$. Near the singularity at $x=1$, it behaves as 
    $\pi (2 |x-1|)^{-1/2}$.} 
\end{figure}

Since we are only interested in  the $l=0$ mode, we drop this  label to simplify the notation. 
In order to obtain the explicit form of the integral equation we must first evaluate the integrals involved in Eq.~(\ref{eq:kinlin}). 
Some details of these calculations are given in  Appendix~\ref{sec:fredholm}. One finds  
\begin{equation}
\label{eq:intV}
\tau \partial_t \mathcal{A}(p,t) = \gamma_N \mathcal{A}(p,t) + 
\int_0^\infty \mathcal{V}\left(\frac{p}{p'}\right)\mathcal{A}(p',t) 
\frac{d p'}{p'} , 
\end{equation}
where the dimensionless constant $\gamma_N$ and $\tau$ with dimension of time are given in (\ref{eq:gammaN}) and (\ref{eq:tau}), respectively. 
The kernel $\mathcal{V}(x)$ is given in Eqs.~(\ref{eq:VV}), (\ref{eq:RR}),  (\ref{eq:SS}) and is depicted 
in Fig.~\ref{fig:kernel}.
This equation is a special case of the general equation which arises in the linearization of the kinetic equation 
about the Kolmogorov spectrum.  The resulting equation has the homogeneous form~\cite{book}  
\begin{equation}
 \partial_t \mathcal{A}(p,t) = 
 p^{-h} \int_0^\infty \mathcal{U}\left(\frac{p}{p'}\right)\mathcal{A}(p',t)  \frac{d p'}{p'} .
\end{equation} 
In the present case, $h=0$ and $\mathcal{U}(x)$ is expressed  as $\gamma_N \delta(x-1) + \mathcal{V}(x)$. 

\section{\label{sec:stb} The Mellin function and the asymptotic behavior of the solution}

The Mellin transform is a very useful tool for solving the integral equation  (\ref{eq:intV}). In terms of the Mellin transform  of 
$\mathcal{V}$, we define 
\begin{equation}
W(s) \equiv  \gamma_N + \int_0^\infty  \mathcal{V}(x)  x^{s-1} d x = \gamma_N+W_\mathcal{V}(s), 
\end{equation}
and thus 
the Mellin image $\mathcal{F}(s,t)$  of $\mathcal{A}(p,t)$ is a solution to 
\begin{equation}
\tau \partial_t \mathcal{F}(s,t) =  W(s) \mathcal{F}(s,t).  
\end{equation} 
Hereafter we will use $t$ and $p$ to denote the dimensionless variables  $t/\tau$ and $p \mathit{l}$, where 
 $\mathit{l}^{-1}$ is some  momentum scale entering into the initial condition $\mathcal{A}(p, t=0)$. 
With these notation, 
the solution of the integral equation may be written as  
\begin{equation}
\label{eq:solA}
\mathcal{A}(p,t) =  \frac{1}{2\pi i} \int_{\sigma - i \infty}^{\sigma+i \infty} p^{-s} 
\Phi(s) \exp \left(W(s) t \right) d s , 
\end{equation}
where $\Phi(s)$  is the Mellin transform of the initial condition 
$\mathcal{A}(p, t=0)$
\begin{equation}
\Phi(s) \equiv \mathcal{F}(s ,t=0)=\int_0^\infty  \mathcal{A}(p, t=0)\,  p^{s-1} d p ,
\end{equation} 
and  $\sigma$  stands somewhere in $-2 < \sigma < 5$, corresponding to the fundamental strip where $W(s)$ is analytic.  
This property follows from the behavior of $\mathcal{V}(x)$
\begin{eqnarray}
\mathcal{V}(x) \sim -4 x^2 \left( \ln x - \frac{2}{3}+\ldots\right), \quad & x \rightarrow 0, \\
\mathcal{V}(x) \sim    \frac{4}{x^5} \left( \ln x + \frac{2}{3}+\ldots\right), \quad & x \rightarrow \infty. 
\end{eqnarray}
which shows that the Mellin function has double poles at $s=-2$ and $s=5$,  
\begin{eqnarray}
\label{eq:as2}
W(s) \sim \frac{4}{(s+2)^2}, \qquad s \rightarrow -2, \\
\label{eq:as3}
W(s) \sim \frac{4}{(s-5)^2}, \qquad s \rightarrow 5. 
\end{eqnarray} 

\begin{figure}
\centering
    \includegraphics{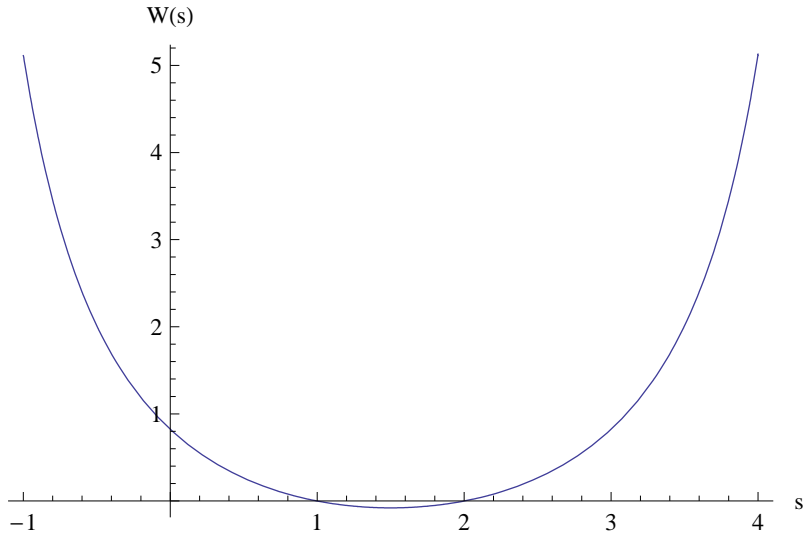}
    \caption{\label{fig:mellin}(color online). A plot of the Mellin function $W(s)$ for real values of $s$. The symmetry with respect to $s=3/2$ is clear.
    The intercept is $W(0) \approx  0.8225$. }
\end{figure}

Some indications about the computation of the Mellin function are given in appendix \ref{sec:mellin}. 
One finds that  $W(s)$  has  zeros at $s=1$ and $s=2$ (see Fig.~\ref{fig:mellin}),  
with the corresponding  solutions $\mathcal{A}=p^{-1}, p^{-2}$. 
Another property is the symmetry with respect to the line $\text{Re} \,s = 3/2$, 
\begin{equation}
W(3/2+s) = W(3/2 - s). 
\end{equation}
The Taylor series about $s=3/2$ reads 
\begin{equation}
W(s) \sim 
-0.0786 +  0.3061 (s-3/2)^2, \qquad  s \rightarrow 3/2 .
\end{equation}

\subsection{The asymptotics as $ t^{-1}| \ln p\,|\rightarrow +\infty$}
We begin the study of the stability by deriving from (\ref{eq:solA}) the infrared behavior of an evolved initial perturbation for arbitrary $t$.
We require that $\sigma < 1$ for the integral  
corresponding to the departure of the particle density 
\begin{equation}
\label{eq:deltan}
\delta n(t) =   \int_0^\infty n^0(p) \mathcal{A}(p,t)\, \frac{4\pi p^2 d p}{(2\pi)^3} 
\end{equation}
to converge as  $p \rightarrow 0$.
To evaluate (\ref{eq:solA}), we apply the steepest descent method to the integral 
\begin{equation}
\label{eq:intx}
\mathcal{A}(p,t) =  \frac{1}{2\pi i} \int_C \Phi(s)   \exp \left(x s  + W(s) t \right) d s , 
\end{equation} 
where $x\equiv-\ln p$  with  $t^{-1}\, x \rightarrow \infty$,  and $C$ is a contour  which runs along a straight line parallel to the imaginary-$s$ axis, 
with $-2<  \text{Re}\,s < 1$.
We assume that the initial perturbation has compact support, and 
its Mellin image vanishes sufficiently rapid as $|\text{Im} s| \rightarrow \infty$
to permit the existence of this integral. 
Note in passing that this property enables us to put the upper limit of the integral  (\ref{eq:deltan}) as infinity. 
The saddle point occurs when $W'(s) =  -x/t$.
To find this location as $t^{-1}\, x \rightarrow \infty$  we may use the asymptotic estimation (\ref{eq:as2}) and obtain
\begin{equation}
s = -2 + 2 \left( \frac{t}{x}\right)^{1/3} ,   
\end{equation}
which depends on $x/t$. 
A change of variable  $s = -2 + 2 (t/x)^{1/3} u$ fixes the movable saddle point at $u=1$, and the 
integral (\ref{eq:intx})  can be approximated by 
\begin{equation}
\frac{2 t^{1/3}  e^{-2 x}}{x^{1/3}}  \int_{1-i  \infty}^{1+i  \infty} \tilde{\Phi}(u)
\exp\left( x^{2/3} t^{1/3} (2 u +1/u^2) \right) \frac{d u}{2\pi i} ,
\end{equation}
where $\tilde{\Phi}(u) = \Phi(s(u))$.  
Now the steepest descent method directly produces  the asymptotics
\begin{equation}
\label{eq:finp}
\mathcal{A}(p,t) \sim  \frac{\Phi(-2)}{\sqrt{3\pi}}  \frac{p^2 t^{1/6}}{(-\ln p)^{2/3} }
\exp{\left( 3 t^{1/3} (-\ln p)^{2/3} \right)} , \quad t^{-1}\, \ln p \rightarrow -\infty. 
\end{equation}
This result shows that, as $t\rightarrow +\infty$,  
the initial perturbation becomes very large in the region where $p>0$ and $t^{-1}\ln p \rightarrow -\infty$,
but preserves the infrared convergence of the integral for the particle density. 

A similar estimation  using the steepest decent contour in the neighborhood of 
$s=5 - 2(t/\ln p)^{1/3}$  
gives the leading behavior as $t^{-1}\, \ln p \rightarrow \infty$:
\begin{equation}
\label{eq:finpdos}
\mathcal{A}(p,t) \sim  \frac{\Phi(5)}{\sqrt{3\pi}}  \frac{ t^{1/6}}{p^5 (\ln p)^{2/3} }
\exp{\left( 3 t^{1/3} (\ln p)^{2/3} \right)} , \quad t^{-1}\, \ln p \rightarrow \infty. 
\end{equation}
where now $\sigma > 1$ in Eq.~(\ref{eq:solA}), for the integral 
\begin{equation}
   \int_0^\infty n^0(p) \mathcal{A}(p,t)\, \frac{4\pi p^2 d p}{(2\pi)^3} 
\end{equation}
to converge as  $p \rightarrow \infty$.
 
\subsection{The asymptotics as $t^{-1} | \ln p\, | \rightarrow 0$}

In this regime,  the relevant saddle point occurs near $s_0 = 3/2$ at 
\begin{equation}
 s_0 \sim \frac{3}{2} + \frac{ \ln p}{t\, W''(3/2)} ,  \qquad  t^{-1} \ln p  \rightarrow 0 . 
\end{equation}
Thus, the steepest descent evaluation yields the exponentially decreasing asymptotics 
\begin{equation}
\label{eq:tasym}
\mathcal{A}(p,t) \sim \frac{\Phi(3/2)}{p^{3/2} \sqrt{2 \pi W''(3/2) t }} \, \exp \left(W(3/2) t \right), \qquad  t^{-1} \ln p  \rightarrow 0. 
\end{equation}
The behavior of the function $\mathcal{A}(p,t)$ in this regime bears some resemblance with what is called interval instability  in \cite{book}. Notice nevertheless that the instability results in \cite{book} can not be directly applied in our case since $h=0$. 


\subsection{The asymptotics of  $\delta n(t; p_0)$} 

A more conclusive indication about the nature of the instability can be easily obtained by studying the induced change 
of the particle number density $\delta n(t; p_0)$ corresponding to the modes within the sphere $p < p_0$. 
Here $p_0$ refers to an arbitrary scale which is smaller than the cut-off scale  corresponding to the critical temperature. 
Let us consider (\ref{eq:deltan}) with a finite upper limit  of integration $p_0$. A first integration over $p$ gives 
\begin{equation}
\delta n(t; p_0) = 
\frac{m T }{\pi^2} \int_C \frac{p_0^{1-s}}{1-s} \Phi(s) e^{t W(s)} 
\frac{d s}{2\pi i}, \qquad -2< \text{Re}\, s < 1, 
\end{equation}
where $C$  runs parallel to the imaginary-$s$  axis. 
Since the integrand falls off sufficiently rapidly at large $|\text{Im}\, s|$, 
the theorem of residues yields  the exact expression
\begin{equation}
\label{eq:exact}
\delta n(t; p_0) = \frac{m T }{\pi^2}  \Phi(1) e^{t W(1)}+ 
\frac{m T }{\pi^2} \int_{C'} \frac{p_0^{1-s}}{1-s} \Phi(s) e^{t W(s)}
\frac{d s}{2\pi i}, 
\end{equation}
where the new contour $C'$ runs parallel to the imaginary-$s$  axis with $1 < \text {Re}\,s < 5$.  
The first piece is the contribution of the single pole at $s=1$, which has been encircled in a negative, clockwise sense.  
Noting that $W(1)=0$ and making use of 
\begin{equation}
\delta n(0; \infty) = \frac{1 }{2 \pi^2}\int_0^\infty n^0(p) \mathcal{A}(p,t=0)\, p^2 d{p} =  \frac{m T }{\pi^2}  \Phi(1) , 
\end{equation}
the first piece of Eq.~(\ref{eq:exact}) becomes time-independent  and insensitive to the size posed by $p_0$. 
It corresponds exactly to the particle density coming from \emph{all} the modes of the initial perturbation, and 
gives the leading behavior as $t \rightarrow \infty$.
With regard to second term, the steepest descent evaluation along $\text{Re}\, s = 3/2$ yields 
\begin{equation}
 \int_{C'} \frac{p_0^{1-s}}{1-s} \Phi(s) e^{t W(s)} \sim 
-\frac{\Phi(3/2) \sqrt{2} }{\sqrt{\pi W''(3/2) p_0 t} } e^{W(3/2) t}, \qquad  t \rightarrow \infty, 
\end{equation} 
which shows an exponential decay law  since $W(3/2) \approx -0.0786< 0$. 
These results  reveal that an arbitrary initial perturbation with  support  in some interval  to  the right  of $p_0$, 
evolves towards the infrared momentum region. This gives rise to a situation where all the initial excess or defect of particles, depending on the sign of $\delta n(0, \infty)$, appears finally located in the interval $(0, p_0)$, irrespective of  the magnitude of $p_0$. 
This is rather peculiar and seems to be consistent with some kind of generation of a large-scale structure.

\section{\label{sec:end}Conclusion}

In this paper  we have considered the problem of the stability of the Rayleigh-Jeans equilibrium solution of the theory of weak wave turbulence,  
in the case of the four-wave contact interactions with a non-relativistic dispersion law. 
It is believed that this theory captures the basics facts of the 
non-equililibrium processes  which occur in  weakly interacting Bose gases 
in the `classical' regime  where the occupation number is large. 

Based on the general theory of the stability of weak-turbulence  Kolmogorov spectra  
presented in Ref.~\cite{book},
we have first derived the explicit expression for  the linear evolution equation for the 
scalar ($l=0$) perturbations, and  then  we have applied the method of Mellin function to analyze the stability of 
solutions. 
Had the index $h$ been positive, the sufficient condition of Balk and Zakharov,  $W(0)>0$, 
would immediately imply the instability of the initial perturbation in the sense of Ref.~\cite{book}. 
However, this effective criterion for  checking the stability is inapplicable in the present case. 
In order to see the the character of  the evolution of an arbitrary initial perturbation, 
we have evaluated  the leading asymptotic behavior  
as  $|\ln p\,| \rightarrow \infty$
with $t$ held fixed, and as  $t\rightarrow \infty$ with $p$ held fixed. 
But a better understanding about the time evolution of a generic solution comes from 
the asymptotics  for the integrated particle number as $t\rightarrow \infty$. 
This evaluation shows clearly the main feature of the evolution of the perturbation: 
the initial departure of the particle density concentrates as $t\rightarrow \infty$ into an 
interval of arbitrarily small width around the origin in momentum space. 
Let us remark again that numerical work based on the nonlinear 
kinetic equation~(\ref{eq:kin4}) strongly indicates  
a finite time  for the formation of  a singularity, 
according to the scenario described in Refs.~\cite{Semikoz1,Semikoz2,Josserand,Lacaze} 
where the nonlinear effects are important.   
We must then be very cautious in seeing the asymptotic instability of the linearized equation around a Rayleigh-Jeans spectrum 
as a hint of the Bose-Einstein condensation. 

Finally, it is worthy mentioning that natural extensions of this work include the analysis of the stability of the Rayleigh-Jeans spectrum 
of  three-wave kinetic equations such as those that arise at the final stage of the condensation process, and the stability for $T>T_c$ where the chemical potential is finite. Work in this direction is in progress.

\section*{Acknowledgments}

The work of M. E. is supported by the Spanish MICINN under Grant  MTM2008-03541 and by the Basque Government under Grant No. IT-305-07.
The work of M. A. V. is partially  supported by 
the Spanish Consolider-Ingenio 2010 Programme CPAN (CSD2007-00042) and by 
the Basque Government under Grant No. IT-357-07.


\appendix

\section{Reduction to Fredholm form}
\label{sec:fredholm}

Here we give some details about the reduction of the linearized kinetic equation for $l=0$ to Fredholm form. First we compute the constant $N$. The angular integration over the orientation of $\bm{p}_3$ in the terms of the integrand of (\ref{eq:kinlin}) containing the functions $A(\bm{p}_1,t)$ and  $A(\bm{p}_2,t)$ may be expressed in a series of Legendre polynomials of 
$\cos\theta_{12}= \hat{\bm{p}}_1 \cdot\hat{\bm{p}}_2$
\begin{eqnarray}
\int \delta(\bm{p}_1&+&\bm{p}_2 -\bm{p}_3 -\bm{p}_4) 
  \delta(\omega_1+ \omega_2-\omega_3-\omega_4) d^3 \bm{p}_4 
  d\Omega_{\bm{p}_3} \nonumber \\ 
  &&= \frac{2\pi m}{p_3|\bm{p}_1+\bm{p}_2|} \Theta\left(1 - 
  \left(\frac{p_3^2+\bm{p}_1 \cdot\bm{p}_2}{p_3|\bm{p}_1+\bm{p}_2|}\right)^2 \right) 
  \equiv \sum_{l=0} f_l P_l(\cos\theta_{12}) ,
\end{eqnarray}
where $\Theta$ denotes the Heaviside step function. Only the $f_0$ coefficient will be needed in order to compute $N$ and to derive the integral equation for $\mathcal{A}_0$ By using the standard formula
\begin{equation}
f_l = \frac{2l+1}{2} \int_{-1}^1 f(u) P_l(u), 
\end{equation}
for the coefficients in the expansion $f(u) = \sum_l f_l P_l(u)$, one finds for $p_1<p_2$ 
\begin{eqnarray}
f_0^{(p_1<p_2)}(p_3) &=& \frac{2\pi m}{p_1 p_2} \Theta(p_1-p_3) + 
          \frac{2\pi m}{p_2 p_3} \left[\Theta(p_3-p_1)-\Theta(p_3-p_2)\right] 
          \nonumber \\ 
 &&+\frac{2\pi m \sqrt{p_1^2+p_2^2-p_3^2}}{p_1 p_2 p_3} \left[\Theta(p_3-p_2)-\Theta(p_3-\sqrt{p_1^2+p_2^2} \,)\right]  ,
\end{eqnarray}  
and for $p_1>p_2$ 
\begin{eqnarray}
f_0^{(p_1>p_2)}(p_3) &=& \frac{2\pi m}{p_1 p_2} \Theta(p_2-p_3) + 
          \frac{2\pi m}{p_1 p_3} \left[\Theta(p_3-p_2)-\Theta(p_3-p_1)\right] 
          \nonumber \\ 
 &&+\frac{2\pi m \sqrt{p_1^2+p_2^2-p_3^2}}{p_1 p_2 p_3} \left[\Theta(p_3-p_1)-\Theta(p_3-\sqrt{p_1^2+p_2^2} \,)\right] .
\end{eqnarray}
Note that $f_0(p_3)$ vanishes for $p_3 >\sqrt{p_1^2+p_2^2}$.
To derive the asymptotic relation (\ref{eq:asymp2}) we may insert the expression of 
$f_0(p_3)$ which produces 
\begin{equation}
\int_0^\infty n^0(p_2) n^0(p_3)f_0(p_2) p_3^2\,d p_3 
     \sim -\frac{8\pi m^3 T^2}{p_2^3} \left(\ln\frac{p_1}{p_2}-1\right),\quad 
     p_2 \rightarrow \infty. 
\end{equation} 

With the notation $\tilde{p}_4 = \sqrt{p_1^2+p_2^2-p_3^2}$, the integral over 
$p_3$ affecting $\mathcal{A}_0(p_1,t)$ adopts the form   
\begin{equation}
\label{eq:QQ}
\int_0^\infty \left[-n^0(p_2) n^0(p_3)-n^0(p_2) n^0(\tilde{p}_4)+
     n^0(p_3) n^0(\tilde{p}_4)\right] f_0(p_3) p_3^2\, d p_3  
     =\frac{8\pi m^3 T^2}{p_2^3} Q\left(\frac{p_1}{p_2}\right) ,
\end{equation}
where
\begin{eqnarray}
Q(x)&=&\left[\frac{2x^2 \ln x}{1+x^2} -
 \frac{2 x}{\sqrt{1+x^2}} \text{arctanh} \left(\frac{x}{\sqrt{1+x^2}}\right)\right]
 \Theta(1-x)
 \nonumber \\ 
 &&-\left[\frac{2 x \ln x}{1+x^2} +
 \frac{2 x}{\sqrt{1+x^2}} \text{arctanh} \left(\frac{1}{\sqrt{1+x^2}}\right)\right]
 \Theta(x-1) .
\end{eqnarray}
Thus the resulting three-dimensional $\bm{p}_2$ integral giving $N$ is independent of $p_1$. 
The explicit computation shows that the radial integration of Eq.(\ref{eq:QQ}) is given by
\begin{eqnarray}
\label{eq:gammaN}
\gamma_N \equiv \int_0^\infty Q(x) \frac{d x}{x } &=& -2G - \frac{13\pi^2}{24} -
\frac{1}{2}\ln\left(3-2\sqrt{2}\right)\ln\left(3+2\sqrt{2}\right) \nonumber \\
&&-\frac{1}{16}\ln^2\left(17+12\sqrt{2}\right) 
- \textrm{Li}_2\left(3-2\sqrt{2}\right)+4\textrm{Li}_2\left(-1+\sqrt{2}\right)  \nonumber \\
&\approx&-4.71057, 
\end{eqnarray}
where $G$ is the Catalan's constant and $\textrm{Li}_2$ denotes the dilogarithmic function. Therefore, putting all factors together, we obtain the value of the constant $N$
\begin{equation}
N=\frac{32\gamma_N}{\pi}m a^2 T^2 , 
\end{equation}
with dimension of 1/time. 

A similar integration for the term containing $\mathcal{A}_0(p_2,t)$ produces 
\begin{equation}
-\int_0^\infty n^0(p_3) n^0(\tilde{p}_4)f_0(p_3) p_3^2\, d p_3 
     =\frac{8\pi m^3 T^2}{p_2^3} R\left(\frac{p_1}{p_2}\right) ,
\end{equation}
where
\begin{eqnarray}
\label{eq:RR}
R(x)&=&\left[\frac{2\ln x}{1+x^2} -
 \frac{2}{x \sqrt{1+x^2}} \text{arctanh} \left(\frac{x}{\sqrt{1+x^2}}\right)\right]
 \Theta(1-x)
 \nonumber \\ 
 &&-\left[\frac{2\ln x}{x(1+x^2)} +
 \frac{2}{x\sqrt{1+x^2}} \text{arctanh} \left(\frac{1}{\sqrt{1+x^2}}\right)\right]
 \Theta(x-1).
\end{eqnarray}

The same procedure may be followed to evaluate the contribution of the coefficient of $\mathcal{A}_0(p_3,t)$ in the integrand of (\ref{eq:kinlin}). Now the appropriate 
expansion in Legendre polynomials reads 
\begin{eqnarray}
\int \delta(\bm{p}_1&+&\bm{p}_2 -\bm{p}_3 -\bm{p}_4) 
  \delta(\omega_1+ \omega_2-\omega_3-\omega_4) d^3 \bm{p}_4 
  d\Omega_{\bm{p}_2} \nonumber \\ 
  &&= \frac{2\pi m}{p_2|\bm{p}_1-\bm{p}_3|} \Theta\left(1 - 
  \left(\frac{-p_3^2+\bm{p}_1 \cdot\bm{p}_3}{p_2|\bm{p}_1-\bm{p}_3|}\right)^2 \right) 
  \equiv \sum_{l=0} g_l P_l(\cos\theta_{13}) ,
\end{eqnarray}
where
\begin{eqnarray}
g_0^{(p_1<p_3)}(p_2)&=&\frac{2\pi m \sqrt{p_1^2+p_2^2-p_3^2}}{p_1 p_2 p_3} \left[\Theta(p_2-\sqrt{p_3^2-p_1^2}\,)-\Theta(p_2-p_3)\right] \nonumber \\ 
&&+\frac{2\pi m}{p_2 p_3} \Theta(p_2-p_3) , \\ 
g_0^{(p_1>p_3)}(p_2)&=&\frac{2\pi m}{p_1 p_3} \Theta(p_3-p_2) + 
  \frac{2\pi m}{p_1 p_2} \Theta(p_2-p_3) .
\end{eqnarray}
The asymptotic behavior of $g_0(p_2)$ is
\begin{equation}
g_0(p_2) \sim \frac{2\pi m}{p_2 p_3} \Theta(p_3-p_1) + 
              \frac{2\pi m}{p_1 p_2} \Theta(p_1-p_3), \qquad p_2 \rightarrow \infty, 
\end{equation}
but it is not dangerous for the remaining $p_2$ integration of 
$n^0(p_2)n^0(\tilde{p}_4)$, which is seen to be 
\begin{equation}
2\int_0^\infty n^0(p_2) n^0(\tilde{p}_4)g_0(p_2) p_2^2\, d p_2 
     =\frac{8\pi m^3 T^2}{p_3^3} S\left(\frac{p_1}{p_3}\right) ,
\end{equation}
where
\begin{eqnarray}
\label{eq:SS}
S(x)&=&\left[-\frac{2\ln x}{1-x^2}+
 \frac{2}{x\sqrt{1-x^2}}\arctan\left(\frac{x}{\sqrt{1-x^2}}\right)\right]\Theta(1-x)
 \nonumber \\
 &&+\left[\frac{2\ln x}{x(x^2-1)}+
 \frac{2}{x\sqrt{x^2-1}}\arctan\left(\frac{1}{\sqrt{x^2-1}}\right)\right]\Theta(x-1). 
 \end{eqnarray}
 The behavior 
 \begin{equation}
 S(x) \approx -2\ln x + 2, \qquad x \rightarrow 0, 
 \end{equation}
 determines the asymptotics which has been used in Eq.(\ref{eq:asymp1}). 
 
 Assembling the various contributions produces the integral equation
 \begin{equation}
 \tau \partial_t\mathcal{A}(p,t)=\gamma_N \mathcal{A}(p,t) + 
 \int_0^\infty \mathcal{V}\left(\frac{p}{p'}\right)\mathcal{A}(p',t) 
 \frac{d p'}{p'} , 
 \end{equation}
 where $\mathcal{V}(x)$ is given by
 \begin{equation}
 \label{eq:VV}
 \mathcal{V}(x)=R(x)+S(x), 
 \end{equation}
 and the constant $\tau$ is 
 \begin{equation}
 \label{eq:tau}
 \tau = \frac{\pi}{32 m a^2 T^2} .
 \end{equation}
 Near $x=1$ the kernel exhibits the behavior
 \begin{equation}
 \mathcal{V}(x) \sim \frac{\pi}{\sqrt{2 |x-1|}},\qquad  x \rightarrow 1.
 \end{equation}
The kernel has the properties
\begin{eqnarray}
\int_0^\infty \mathcal{V}(x) \frac{d x}{x} &=& 2G+ \frac{3\pi^2}{8} \approx 5.5330, \\ 
\label{eq:zm}
\int_0^\infty \mathcal{V}(x)\, x \,d x &=& -\gamma_N, 
\end{eqnarray} 
the second assuring that $\mathcal{A}_0 \propto p^{-2}$ is a zero mode of $L$. 
This completes the derivation of the explicit expression for the integral equation.


\section{Computation of the Mellin function} 
\label{sec:mellin}

The integrals involved in the computation of the Mellin function containing  logarithmic terms  
can be expressed in closed form  in terms of the polygamma function $\psi^{(n)}(z)$ for $n=1$:
\begin{eqnarray}
\int_0^1 \frac{x^\alpha \ln x}{1+x^2}\,d x &=&  \frac{1}{16} \psi'\left(\frac{3+\alpha}{4} \right) - \frac{1}{16} \psi'\left(\frac{1+\alpha}{4} \right) , \qquad 
  \text{Re}\, \alpha > -1,  \\ 
\int_0^1 \frac{x^\alpha \ln x}{1-x^2}\,d x &=&  -\frac{1}{4} \psi'\left(\frac{1+\alpha}{2} \right), \qquad 
  \text{Re}\, \alpha > -1.  
\end{eqnarray}
To perform the remainder integration involving the inverse trigonometric functions,  
it is convenient  to use the following series expansion  
\begin{equation}
\arctan u =  \frac{u}{\sqrt{1+u^2}} \sum_{n=0}^\infty \frac{(2n)!}{2^{2n} (n!)^2 (2n+1)} \frac{u^{2n}}{(1+u^2)^n} , 
\quad u^2 < \infty,  
\end{equation}
and its counterpart for $\text{arctanh}\,u$. 
The integration term by term of the resulting series and the subsequent summation 
produces  some combination of  hypergeometric functions 
${}_3 F_2$  and  ${}_2 F_1$ evaluated at $\pm1$. 
This procedure yields  
\begin{eqnarray}
\int_0^1\frac{x^{s-1}}{x \sqrt{1-x^2}} \arctan\left(\frac{x}{\sqrt{x^2-1}} \right)  d x  &=&
\frac{\sqrt{\pi}}{2} \frac{\Gamma\left(s/2\right)}{\Gamma\left(1/2 +s/2\right)} \nonumber \\
&&\times  \, 
{}_3 F_2\left(\frac{1}{2}, \frac{1}{2}, \frac{s}{2}; \frac{3}{2}, \frac{1+s}{2}; 1 \right),  \\  
\int_0^1 \frac{x^{s-1}}{x \sqrt{1+x^2}} \text{arctanh} \left(\frac{x}{\sqrt{1+x^2}} \right)  d x   &=& 
\frac{1}{s(1-s)}\, {}_3 F_2\left(1, \frac{s}{2}, \frac{s}{2}; \frac{1+s}{2}, 1+\frac{s}{2}; -1 \right)  \nonumber \\ 
&& + \frac{\sqrt{2} \ln\left(1+\sqrt{2}\right) }{s-1} \, {}_2 F_1\left(1, \frac{s}{2};  \frac{1+s}{2}; -1 \right) .
\end{eqnarray}
The other integrals in the interval $[1,\infty]$ are obtained from these by making  the change of variable $x \rightarrow 1/x$. 
Assembling the various pieces,  one finds the integral for the Mellin transform of $\mathcal{V}(x)$ 
\begin{eqnarray}
W_\mathcal{V}(s) &=& \int_0^\infty \mathcal{V}(x) x^{s-1} d x \nonumber \\ 
&=&-\frac{2 \sqrt{2} \log \left(1+\sqrt{2}\right)}{s-1}\, {}_2 F_1\left(1, \frac{s}{2};  \frac{1+s}{2}; -1 \right) \nonumber \\
 && +\frac{2 \sqrt{2} \log \left(1+\sqrt{2}\right)}{s-2}\, {}_2 F_1\left(1, \frac{3-s}{2};  \frac{4-s}{2}; -1 \right) \nonumber \\
  &&+\frac{\sqrt{\pi }\, \Gamma \left(\frac{s}{2}\right)}{\Gamma \left(\frac{s+1}{2}\right)}\,
   {}_3 F_2\left(\frac{1}{2}, \frac{1}{2}, \frac{s}{2}; \frac{3}{2}, \frac{1+s}{2}; 1 \right) \nonumber \\ 
   &&+\frac{\sqrt{\pi }\, \Gamma \left(\frac{3-s}{2} \right)}{\Gamma \left(2-\frac{s}{2}\right)} \,
   {}_3 F_2\left(\frac{1}{2}, \frac{1}{2}, \frac{3-s}{2}; \frac{3}{2}, 2-\frac{s}{2}; 1 \right) \nonumber \\ 
   && + \frac{2}{(s-1) s}\, {}_3 F_2\left(1, \frac{s}{2}, \frac{s}{2}; \frac{1+s}{2}, 1+\frac{s}{2}; -1 \right)  \nonumber \\  
    &&+\frac{2}{(s-3) (s-2)} \, {}_3 F_2\left(1, \frac{3-s}{2}, \frac{3-s}{2}; 2-\frac{s}{2}, \frac{5-s}{2}; -1 \right)  \nonumber \\
    &&+\frac{1}{4} \psi' \left(\frac{3-s}{2}\right) -\frac{1}{8} \psi' \left(\frac{3-s}{4}\right) 
   -\frac{1}{4} \psi' \left(1-\frac{s}{2}\right) \nonumber \\ 
    &&+\frac{1}{8} \psi' \left(\frac{5-s}{4}\right) + \frac{1}{4} \psi' \left(\frac{s-2}{4}\right) 
   +\psi' \left(2-s\right) -\frac{4}{(s-2)^2}  , 
\end{eqnarray}
which converges in the strip $-2 < \text{Re}\, s < 5$.


\end{document}